\documentclass{aa}
\usepackage{graphicx}
\begin{document}
   	\title{S\,Ori J053825.4$-$024241: \\ 
	A Classical T Tauri-like object at the substellar boundary}

   	\author{J. A. Caballero\inst{1}
          	\and
   	  	E. L. Mart\'{\i}n\inst{1,2}
          	\and
          	M. R. Zapatero Osorio\inst{3}
          	\and
   	  	V. J. S. B\'ejar\inst{4}
          	\and
   	  	R. Rebolo\inst{1,5}
		\and
   	  	Ya. Pavlenko\inst{6,7}
		\and
   	  	R. Wainscoat\inst{8}
		}


   	\institute{Instituto de Astrof\'{\i}sica de Canarias, E-38205 La
	Laguna, Tenerife, Spain\\
        \email{zvezda@ll.iac.es}
        \and
        University of Central Florida, Dept. of Physics, P.O. Box 162385,
	Orlando, FL 32816-2385, USA 
        \and
        LAEFF-INTA, P.O. Box 50727, E-28080, Madrid, Spain
        \and
        Proyecto Gran Telescopio Canarias, Instituto de Astrof\'{\i}sica de
	Canarias 
	\and
        Consejo Superior de Investigaciones Cient\'{\i}ficas, Spain
	\and
        Centre for Astrophysics Research, University of Hertfordshire, College
	Lane, Hatfield, Hertfordshiere AL10 9AB, United Kingdom  
	\and
        Main Astronomical Observatory of Academy of Sciences of Ukraine,
	Golosiiv Woods, Kyiv-127, 03680, Ukraine 
        \and
        Institute for Astronomy, University of Hawai'i, 2680 Woodlawn Drive, 
	Honolulu, HI 96822, USA       
}

   	\date{Received May 18, 2005; accepted August 08, 2005}

        \abstract{We present a spectrophotometric analysis of S\,Ori
        J053825.4$-$024241, a candidate member close to the substellar
        boundary of the young (1--8\,Myr), nearby ($\sim$360\,pc)
        $\sigma$ Orionis star cluster. Our optical and near-infrared
        photometry and low-resolution spectroscopy
        indicate that S\,Ori J053825.4$-$024241 is a likely cluster
        member with a mass estimated from evolutionary models at
        0.06$^{+0.07}_{-0.02}$\,M$_\odot$, which makes the object
	a probable brown dwarf.  
	The radial velocity of S\,Ori J053825.4$-$024241 is similar 
	to the cluster systemic velocity.
	This target, which we have
        classified as an M\,6.0$\pm$1.0 low-gravity object, shows excess
        emission in the near-infrared and anomalously strong
        photometric variability for its type (from the blue 
        to the $J$ band), suggesting the presence of a surrounding
        disc. The optical spectroscopic observations show a
        continuum excess at short wavelengths and a persistent and
        resolved H$\alpha$ emission (pseudo-equivalent width of
        $\sim-$250\,\AA) in addition to the presence of other
        forbidden and permitted emission lines, which we interpret as
        indicating accretion from the disc and possibly mass loss. We
        conclude that despite the low mass of S\,Ori
        J053825.4$-$024241, this object exhibits some of the
        properties typical of active classical T Tauri stars. 

	\keywords{stars: low mass, brown dwarfs --- 
	stars: individuals (S\,Ori J053825.4$-$024241) ---
	accretion, accretion discs --- 
	stars: variables: others ---
	open clusters and associations: individuals ($\sigma$ Orionis)}}
	
	\titlerunning{A CTT-like object at the substellar boundary}

   	\maketitle
%

\section{Introduction}

T Tauri stars are the bridge between new-born embedded protostars and
zero age main sequence stars. 
The most widely accepted
scenario proposed to explain all their features is the magnetospheric
accretion of mass infalling from a surrounding circumstellar disc.
Bertout, Basri \& Bouvier (1988), Appenzeller \& Mundt (1989) and K\"onigl (1991)
have reviewed the properties of T Tauri stars and their discs.
The existence of a T Tauri-like phase in brown dwarfs, objects below the
hydrogen-burning mass limit ($\sim$0.072\,M$_\odot$ for solar
metallicity) has been suggested in the past few
years.  See, for example, Muzerolle et al. (2003), Barrado y
Navascu\'es \& Mart\'{\i}n (2003), Natta et al. (2004), or the most recent
papers by Mohanty, Jayawardhana \& Basri (2005) and Furlan et al. (2005) and references
therein.  

\object{S\,Ori J053825.4$-$024241} was firstly identified as a
photometric substellar candidate of the young $\sigma$ Orionis cluster
(B\'ejar, Zapatero Osorio \& Rebolo  2004). 
Hence, it is probably 1--8\,Myr old and is
located at a distance of about 360\,pc (see B\'ejar et al$.$ 2001 for
further details on the cluster properties).  
S\,Ori J053825.4$-$024241 exhibits considerable excess emission in the $K_{\rm
s}$-band, whereas the $R-I$ colour appears to be slightly bluer than the colour of
other cluster members of similar magnitude. 
Caballero et al. (2004) found S\,Ori J053825.4$-$024241 to be one of the
most photometric variable sources of its spectral type, with variations as high
as 0.36\,mag in the $I$ band over a few hours.

In this paper we present optical spectroscopy that confirms the
membership of S\,Ori J053825.4$-$024241 in $\sigma$ Orionis at the
borderline between stars and brown dwarfs.  Both photometric and
spectroscopic data suggest that this object is undergoing active mass
accretion. 
The presence of circumstellar discs around low-mass stars and brown dwarfs
in the $\sigma$ Orionis cluster has been inferred by several authors (e.g.,
Oliveira, Jeffries \& van Loon 2004 and Scholz \& Eisl\"offel 2004).
However, the particular peculiarities of S\,Ori J053825.4$-$024241 deserve 
intensive study.

\section{Observations and analysis}

\subsection{Optical spectroscopy} 
\label{optical_spectroscopy}

\begin{figure}
   \centering
   \includegraphics[width=0.48\textwidth]{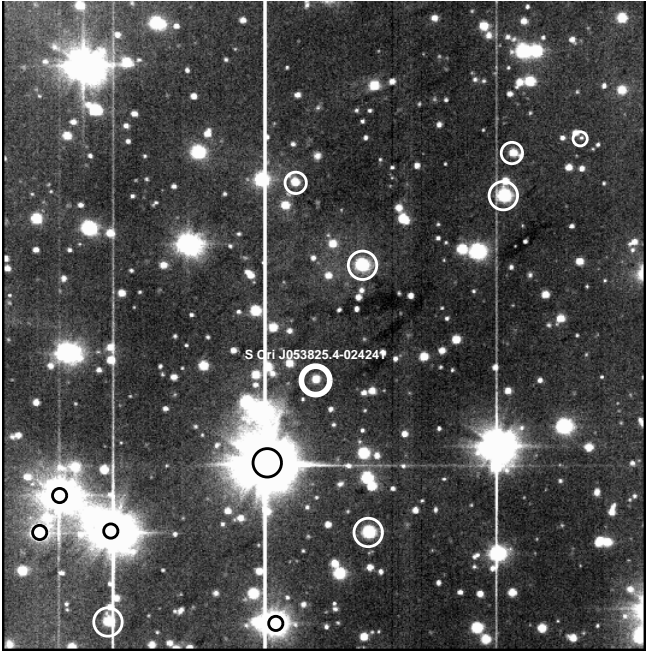}
      \caption{Finder chart of S\,Ori J053825.4$-$024241 (labelled)
      taken in the $I$-band with the IAC-80 telescope. Other stellar
      and substellar members of the $\sigma$ Orionis cluster are
      indicated with circles. Field of view is 6.5 $\times$ 6.5 arcmin$^2$.
      North is at the top and east to the left. }
         \label{image_nIR_iac80}
   \end{figure}

We have obtained low-resolution spectra of S\,Ori
J053825.4$-$024241 using the 2.56 m Nordic Optical Telescope (NOT,
Observatorio del Roque los Muchachos, Spain) and the 10 m Keck I Telescope
(W. M. Keck Observatory, Hawai'i). 
The finder chart of this object is depicted in
Fig.~\ref{image_nIR_iac80}, and Table \ref{log_spec} provides the log
of the spectroscopic observations. Raw data were reduced using tasks within the {\sc
iraf} environment. Reduction steps included bias subtraction,
flat-fielding, optimal extraction, wavelength calibration, and
correction for instrumental response and telluric contribution.

\begin{table}
    \centering
    	\caption[]{Log of the spectroscopic observations}  
        \label{log_spec}
        \begin{tabular}{lcc}
            	\hline
            	\noalign{\smallskip}
		& NOT			& Keck \\
            	\noalign{\smallskip}
            	\hline
            	\noalign{\smallskip}
Date		& 2003 Dec 26		& 2004 Jan 23 \\
MJD		& 52999.93		& 53027.31 \\
No. of spectra	& 3			& 2 \\
Exposure time	& 3 $\times$ 900 s	& 1200 $+$ 390 s \\
Telescope	& NOT			& Keck I \\
Instrument	& {\sc alfosc} 		& {\sc lris} \\
Detector  	& E2V 2k $\times$ 2k 42--40 & Tektronics 2k $\times$ 2k \\
Grism   	& \#5   		& \#4 \\
Blocking filter & GG475   	& GG495 \\
$\lambda$ coverage& 5300--9600\,\AA 	& 5600--8100\,\AA \\
Pixel size	& 0.19 arcsec		& 0.213 arcsec \\
Slit width	& 1.3 arcsec		& 1.0 arcsec \\
Dispersion	& 3.1\,\AA~pixel$^{-1}$	& 1.25\,\AA~pixel$^{-1}$ \\
Resolution      & 12\,\AA               & 4.5\,\AA \\
           	\noalign{\smallskip}
            	\hline
         \end{tabular}
\end{table}

\begin{table*}
    \centering
    	\caption[]{$I$-band magnitude, H$\alpha$ emission, pseudocontinuum
	indices and spectral types from the NOT data}     
        \label{data_spec_alfosc_PC}
        \begin{tabular}{lccccccc}
            	\hline
            	\noalign{\smallskip}
Name			 & $I$$^{\mathrm{a}}$  & pEW (H$\alpha$)       	& PC1			& PC2			& PC3			& PC4			& Spectral type \\
                         & (mag)&     (\AA)    &         		&                    	&               	&               	&               \\
            	\noalign{\smallskip}
            	\hline
            	\noalign{\smallskip}
S\,Ori J053825.4$-$024241 & 16.86 $\pm$ 0.06 & $-$230 $\pm$ 70 		& 1.2 $\pm$ 0.6 	& 1.85 $\pm$ 0.17 	& 1.36 $\pm$ 0.13 	& 1.7 $\pm$ 0.3		& M6.0 $\pm$ 1.0 \\
S\,Ori J053954.3$-$023719 & 17.03 $\pm$ 0.04 & $-$5 $\pm$ 1    		& 1.8 $\pm$ 0.2 	& 2.0 $\pm$ 0.2 	& 1.34 $\pm$ 0.18 	& 1.64 $\pm$ 0.12	& M6.0 $\pm$ 1.0 \\
S\,Ori J053838.6$-$024157 & 16.52 $\pm$ 0.07 & $-$6 $\pm$ 1    		& 1.39 $\pm$ 0.14 	& 1.75 $\pm$ 0.18 	& 1.3 $\pm$ 0.3 	& 1.6 $\pm$ 0.2		& M5.5 $\pm$ 1.0 \\
S\,Ori J053826.1$-$024041 & 16.87 $\pm$ 0.06 & [$-$2,+2]$^{\mathrm{b}}$ & 1.4 $\pm$ 0.4 	& 2.2 $\pm$ 0.5 	& 1.19 $\pm$ 0.18 	& 1.5 $\pm$ 0.3		& M5.0 $\pm$ 2.0$^{\mathrm{c}}$ \\
           	\noalign{\smallskip}
            	\hline
         \end{tabular}
	\begin{list}{}{}
	\item [$^{\mathrm{a}}$] From B\'ejar et al. (2004).
	\item [$^{\mathrm{b}}$] pEW(H$\alpha$) = $-$4 $\pm$ 2\,\AA ~(from
	Barrado y Navascu\'es et al. 2003).
	\item [$^{\mathrm{c}}$] M6.0 spectral type matches better to the 
	spectroscopic cluster sequence.
	It is an M8-type binary dwarf according to Barrado y Navascu\'es et
	al. (2003).
	\end{list}
\end{table*}

We used the Andaluc\'{\i}a Faint Object Spectrograph and Camera ({\sc
alfosc}) instrument attached to the Cassegrain focus of the NOT
to obtain a spectrum of S\,Ori J053825.4$-$024241 (R
$\sim$600).  
See Table \ref{log_spec} for details on the observations and the
instrumental configuration.
We note the strong fringing redwards of 7600\,\AA, and the low number of
photons bluewards of $\sim$6000\,\AA.  Observations were hampered by
poor weather conditions (cirrus and high relative humidity).

\begin{figure}
   \centering
   \includegraphics[width=0.48\textwidth]{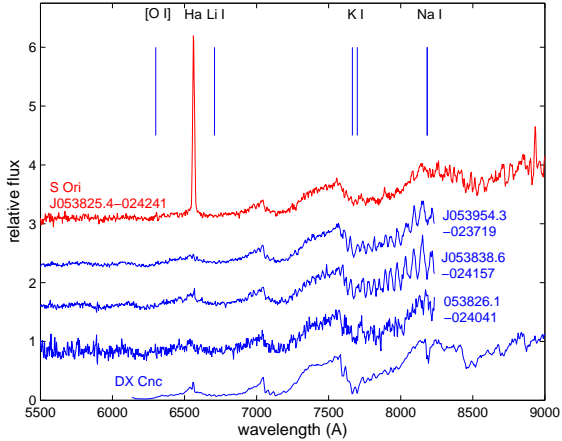}
      \caption{Low-resolution NOT spectra of $\sigma$ Orionis
   candidates. 
   The top spectrum corresponds to S\,Ori J053825.4$-$024241.
   For the other objects, the data redwards of 8200\,\AA~are not displayed
   because of their very poor signal-to-noise ratio. 
   The spectrum of the field M6.5V-type dwarf DX Cnc (Mart\'\i n et al. 1996),
   which has been degraded to the same resolution as the NOT data, is also
   displayed for comparison.  
   Spectra have been shifted in the vertical axis for clarity. 
   Several atomic features are labelled.
   Coloured versions of all the figures can be found in the electronic
   version.}
   \label{alfosc} 
   \end{figure}

We also obtained spectra
of three additional $\sigma$ Orionis candidate members of similar
brightness on the same night ($I$\,$\sim$\,17\,mag, namely \object{S
Ori J053954.3$-$023719}, \object{S\,Ori J053838.6$-$024157} and
\object{S\,Ori J053826.1$-$024041}). These were found in the photometric
searches of B\'ejar et al. (2004) and Caballero et
al. (2004). These authors reported on the short-scale variable
nature of S\,Ori J053826.1$-$024041. Barrado y Navascu\'es et
al. (2003) classified S\,Ori J053826.1$-$024041 as an M8-dwarf member
of the cluster. Exposure times were 3\,$\times$\,900\,s for S\,Ori
J053825.4$-$024241, and 1000\,s for each of the other three objects.
Figure \ref{alfosc} shows the NOT spectra, which were corrected for
telluric lines by observing featureless stars. 

Spectral types were derived by measuring the pseudocontinuum indices (PC) of
Mart\'\i n, Rebolo \& Zapatero Osorio (1996) and by direct comparison to
well-known late-type spectroscopic standard stars. 
We provide our measurements in Table \ref{data_spec_alfosc_PC}. 
We note that the PC1 index is affected by the broad H$\alpha$
emission of S\,Ori J053825.4$-$024241, and that the PC4 index suffers
from the strong fringing of our data. 
PC2 and PC3 remain, however, as reliable indicators of the spectral type. 
In case of strong veiling due to a blue continuum excess, the PC indices
may provide spectral types that are slightly earlier than the real ones. 
The polynomial fitting to the PC index--spectral type relations given in
Mart\'\i n et al. (1996) yields spectral types between M5 and M6 for the four 
objects in our work. In particular, we have determined that the 
spectral type of S\,Ori J053825.4$-$024241, the main target of our 
paper, is M6 with an uncertainty of one subtype. 

As seen from Fig. \ref{alfosc}, the most prominent spectral feature of
S\,Ori J053825.4$-$024241 is H$\alpha$, which is seen in strong
emission. We measured a pseudoequivalent width (pEW, equivalent width
with respect to the observed continuum) of $-$230$\pm$70\,\AA~on the
NOT spectrum, which makes S\,Ori J053825.4$-$024241 one of the largest
H$\alpha$ emitters among $\sigma$ Orionis late-type members. The other
three objects do not display such intense H$\alpha$
lines (see Table \ref{data_spec_alfosc_PC}). 
The forbidden line of [O\,{\sc i}] $\lambda$6300.3 is also detected 
in emission with a moderate strength in the
NOT spectrum of S\,Ori J053825.4$-$024241. Regarding
absorption features, all $\sigma$ Orionis objects show rather 
weak K\,{\sc i} $\lambda\lambda$7664.9,7699.0 and Na\,{\sc i}
$\lambda\lambda$8183.3,8148.8 lines as compared to field dwarfs of
similar types (see Fig. \ref{alfosc}). Although the
doublets are blended at the low resolution of the NOT data, they
should appear with pEWs of 20\,\AA~and 6\,\AA~in $\sim$M6 V-type 
spectra, respectively. 
However, we can only impose upper limits that are
significantly smaller. This is indicative of
low-gravity atmospheres (Mart\'\i n et al. 1996; Luhman et al. 1997)
or of the presence of a strong continuum excess, both supporting the
young age of the objects and their membership of the $\sigma$ Orionis
cluster. We note that our pEWs of the Na\,{\sc i}
$\lambda\lambda$8183.3,8148.8 doublet are in agreement with the
measurements of Kenyon et al. (2005) obtained for various M5--6-type
$\sigma$ Orionis members.
Another excellent indicator of youth is the presence of 
Li\,{\sc i} $\lambda$6707.8 absorption. The low resolution and
moderate signal-to-noise ratio of the NOT spectra do not allow a clear
detection of this feature. 
Table \ref{data_spec_alfosc_pEW} provides our
measurements of S\,Ori J053825.4$-$024241. 

\begin{table}
    \centering
    	\caption[]{S\,Ori J053825.4$-$024241: {\sc NOT} data}    
        \label{data_spec_alfosc_pEW}
        \begin{tabular}{lr}
            	\hline
            	\noalign{\smallskip}
Line						& pEW (\AA) \\
            	\noalign{\smallskip}
            	\hline
            	\noalign{\smallskip}
$[$O {\sc i}$]$	$\lambda$6300.3 		& $-$2.5$ \pm$ 1.0 \\	  
H$\alpha$ $\lambda$6562.8	   		& $-$230 $\pm$ 70 \\   
Li {\sc i} $\lambda$6707.8	   		& $< +$0.7 \\	       
K {\sc i} $\lambda$7664.9$^{\mathrm{a}}$	& $< +$2.0 \\	       
K {\sc i} $\lambda$7699.0$^{\mathrm{a}}$	& $< +$2.0 \\
Na {\sc i} $\lambda$8183.3$^{\mathrm{a}}$	& $< +$2.0 \\	       
Na {\sc i} $\lambda$8184.8$^{\mathrm{a}}$	& $< +$2.0 \\
           	\noalign{\smallskip}
            	\hline
         \end{tabular}
	\begin{list}{}{}
	\item [$^{\mathrm{a}}$] Line measurements affected by fringing.
	\end{list}
\end{table}

\begin{figure}
   \centering
   \includegraphics[width=0.48\textwidth]{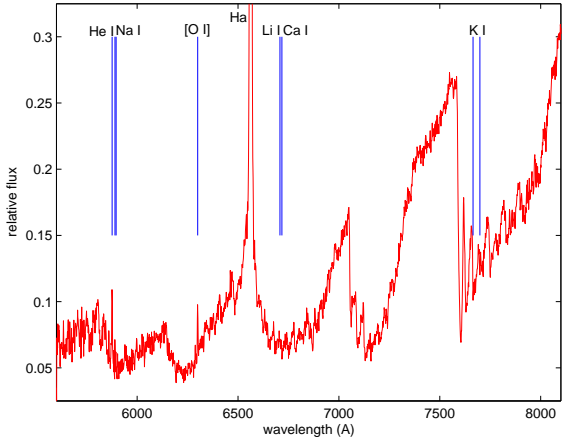}
      \caption{Keck spectrum of S\,Ori J053825.4$-$024241. No
      correction for instrumental response and telluric contribution
      has been applied.}
         \label{lris}
   \end{figure}

\begin{figure}
   \centering
   \includegraphics[width=0.48\textwidth]{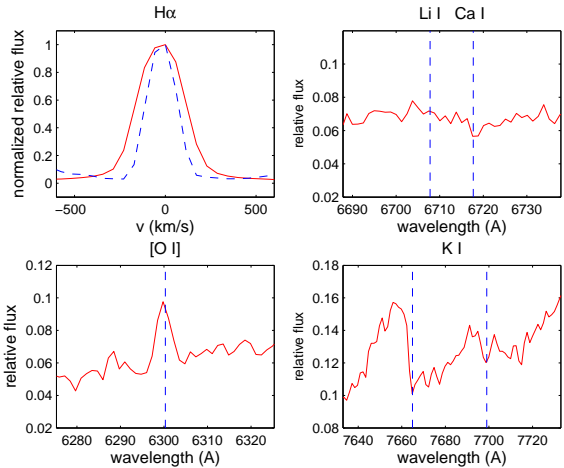}
      \caption{Details of the Keck spectrum of S\,Ori
      J053825.4$-$024241. {\sl (Top left)} The H$\alpha$ profile (full
      line) normalized to the peak and compared to one sky emission
      line (dashed line) at around a similar wavelength. {\sl (Top
      right)} Lithium non-detection with an upper limit on the pEW of
      $+$0.3\,\AA. {\sl (Bottom left)} Emission from [O\,{\sc i}]. {\sl
      (Bottom right)} Spectral region around the K\,{\sc i}
      $\lambda\lambda$7664.9,7699.0 doublet.}
         \label{lris_lines}
   \end{figure}

To check the stability of the H$\alpha$ emission of S\,Ori
J053825.4$-$024241, we collected two optical spectra with a higher
resolution ($R \sim$1700) of 1200\,s and 390\,s about one month later than the 
NOT data (see Table \ref{log_spec}). The Low Resolution Imaging
Spectrograph ({\sc lris}) was used at the Keck I Telescope. 
The Keck spectra were
wavelength-calibrated using emission sky lines (Osterbrock et
al. 1996), which were observed along with the target. The root-mean-square
of the polynomial fit to the calibration is typically 0.28\,\AA. 
The data were not
corrected for instrumental response or terrestrial contribution
because no flux standard star was observed owing to very poor weather
conditions.  Of the two spectra, the first exposure shows better
signal-to-noise ratio. This spectrum is depicted in
Figure \ref{lris}. The second spectrum is of very poor quality. However,
H$\alpha$ is clearly detected in strong emission, as it is in the first
spectrum. 
The Keck data confirm the persistent emission of S
Ori J053825.4$-$024241.
The profile of the observed H$\alpha$ emission is shown in
Figure \ref{lris_lines}. We have plotted relative fluxes as a function
of the velocity shift from line center and scaled to the peak
H$\alpha$ flux of the object. H$\alpha$ appears rather symmetric and
broad. It is resolved as compared to telluric emission lines at
similar wavelengths.


Besides H$\alpha$, we have also detected emission from He\,{\sc i}
(D$_3$) $\lambda$5875.8 and [O {\sc i}] $\lambda$6300.3
(Fig. \ref{lris_lines}). No significant variations are seen between
the two consecutive Keck spectra. Pseudo-equivalent widths measured from the best-quality
Keck spectrum are given in Table \ref{data_spec_lris}. 
As inferred from the strength
of the emission lines (Tables \ref{data_spec_alfosc_pEW} and
\ref{data_spec_lris}), the activity level of S\,Ori J053825.4$-$024241
remains approximately constant between the two epochs of
observation. We also provide in Table \ref{data_spec_lris} the pEWs
of several Na\,{\sc i}, K\,{\sc i}, Li\,{\sc i} and
Ca\,{\sc i} lines. Note that the K\,{\sc i} doublet is severely affected by
telluric absorption. We remark that only upper limits can be set for
Li\,{\sc i} $\lambda$6707.8 (Fig. \ref{lris}) and the Na\,{\sc i}
$\lambda\lambda$5890.0,5895.9 doublet. The Na\,{\sc i} and K\,{\sc i}
lines (Fig. \ref{lris}) are detected much weaker than expected for
high-gravity dwarfs of similar spectral type.

The heliocentric radial velocity of S\,Ori J053825.4$-$024241 was
computed via Fourier cross-correlation of the object best Keck
spectrum with the field standard star \object{V1298 Aql} (vB 10, M8 V,
$v_h$\,=\,35.3\,$\pm$\,1.5\,km\,s$^{-1}$, Tinney \& Reid 1998). This
template star was also observed with the Keck Telescope and similar
instrumentation on a previous run. We took special care in
cross-correlating spectral windows (6620--6840\,\AA, and
7000--7260\,\AA) that are not affected by telluric absorptions and
that contain many photospheric lines (particularly TiO). Additionally,
we considered parts of the spectrum free of emission lines. We derived
$v_h$\,=\,+32\,$\pm$\,13\,\,km\,s$^{-1}$ for S\,Ori
J053825.4$-$024241, which is in full agreement within the error bars
with the mean heliocentric radial velocity of $\sigma$\,Orionis
members (Walter et al. 1998; Zapatero Osorio et al. 2002a). 
Velocities were also computed for each emission line of the spectrum by
measuring line centroids. 
Observed velocities were corrected for the rotation
of the Earth, the motion of the Earth about the Earth--Moon barycentre,
and the orbit of the barycentre about the Sun to yield heliocentric
radial velocities, which are listed in Table \ref{data_spec_lris}. 
We note that the forbidden emission due to
$[$O\,{\sc i}$]$ appears blue-shifted in velocity.

\begin{table}
    \centering
    	\caption[]{S\,Ori J053825.4$-$024241: Keck data}    
        \label{data_spec_lris}
        \begin{tabular}{lccccc}
            	\hline
            	\noalign{\smallskip}
Line				& pEW (\AA)		& $v_h$ (km\,s$^{-1}$) \\
            	\noalign{\smallskip}
            	\hline
            	\noalign{\smallskip}
He {\sc i} $\lambda$5875.8	& $-$3.5 $\pm$ 1.0 	& $+$29 $\pm$ 13$^{\mathrm{a}}$ \\	  
Na {\sc i} $\lambda$5890.0	& $<$ $+$0.1		& --  	\\	  
Na {\sc i} $\lambda$5895.9	& $<$ $+$0.1		& --  	\\	  
$[$O {\sc i}$]$	$\lambda$6300.3	& $-$2.1 $\pm$ 0.5 	& $+$7 $\pm$ 13$^{\mathrm{a}}$	\\	  
H$\alpha$ $\lambda$6562.8 	& $-$260 $\pm$ 30 	& $+$23 $\pm$ 13$^{\mathrm{a}}$ \\	  
$[$N {\sc ii}$]$ $\lambda$6583.5& [$-$0.2, $+$0.2]      & --	\\	  
Li {\sc i} $\lambda$6707.8	& $<$ $+$0.3      	& --	\\	  
Ca {\sc i} $\lambda$6717.7	& $+$0.5 $\pm$ 0.1	& $+$32 $\pm$ 13$^{\mathrm{b}}$ \\
K {\sc i} $\lambda$7699.0	& $+$0.5 $\pm$ 0.1	& $+$32 $\pm$ 13$^{\mathrm{b}}$ \\
           	\noalign{\smallskip}
            	\hline
         \end{tabular}
	\begin{list}{}{}
	\item [$^{\mathrm{a}}$] Velocity obtained from the line centroid.
	\item [$^{\mathrm{b}}$] Velocity obtained from the cross-correlation technique.
	\end{list}
\end{table}

\subsection{Optical photometry}

We have also conducted multiwavelength optical and near-infrared
broad-band observations of S\,Ori J053825.4$-$024241 using the $R$,
$I$, $J$ and $H$ filters and in white light. 
These data, which we have added to previous
measurements available in the literature (2MASS photometry, B\'ejar et
al. 2004, and Caballero et al. 2004, 2005), have been employed to monitor 
photometric variability on different timescales. Broad-band
photometry has been obtained using the following telescopes: the 0.8 m
IAC-80, the 1 m Optical Ground Station (OGS) and the 1.52 m Telescopio 
Carlos S\'anchez (TCS) at the Teide Observatory, the 2.5 m Isaac
Newton Telescope (INT) at Roque de los Muchachos Observatory, the
3.5 m Calar Alto Telescope at Calar Alto Observatory, and the 8 m Very
Large Telescope UT1 ({\it Antu}) at Paranal Observatory. In
Table \ref{log_phot} we provide the log of the observations. Listed are the
Modified Julian Date of the beginning of the observations, total
number of exposures and integration time per exposure, temporal
coverage per observing date, and filter and instrument used. 

Raw frames were reduced within {\sc iraf} using standard techniques in
the optical and near-infrared wavelengths. 
We have conducted a careful differential photometric analysis of all the
data in the same manner as in Caballero et al. (2004).
We refer to this paper for a description of the analysis.

\begin{table*}
    \centering
    	\caption[]{Log of the photometric observations.}  
        \label{log_phot}
        \begin{tabular}{lccccccr}
            	\hline
            	\noalign{\smallskip}
Observing date	& Start		& Exposure 		& $\Delta$t	& Telescope	& Instrument 	& Filter  	& Reference \\	
		& MJD		& time (s)		& (h)		& 		& 	 	& 	  	& \\	
            	\noalign{\smallskip}
            	\hline
            	\noalign{\smallskip}
1998 Jan 22	 & 50835.929	 & 1 $\times$ 1800	   & $<$1	 & IAC-80	 & {\sc ccd}	 & $I$  	 & B\'ejar et al. (2004) \\
1998 Jan 22	 & 50835.983	 & 1 $\times$ 1800	   & $<$1	 & IAC-80	 & {\sc ccd}	 & $R$  	 & B\'ejar et al. (2004) \\
1999 Jan 22	 & 51201.047	 & 1 $\times$ 720	   & $<$1	 & TCS  	 & {\sc cain-2}  & $J$  	 & this paper \\
1999 Jan 24	 & 51202.870	 & 1 $\times$ 720	   & $<$1	 & TCS  	 & {\sc cain-2}  & $J$  	 & B\'ejar et al. (2004) \\
2000 Dec 30	 & 51909.075	 & 5 $\times$ 1500	   & 1.4	 & INT  	 & {\sc wfc}	 & $I$  	 & Caballero et al. (2004) \\
2000 Dec 31	 & 51909.901	 & 10 $\times$ 1500	   & 4.7	 & INT  	 & {\sc wfc}	 & $I$  	 & Caballero et al. (2004) \\
2001 Jan 01	 & 51911.055	 & 6 $\times$ 1500	   & 5.2	 & INT  	 & {\sc wfc}	 & $I$  	 & Caballero et al. (2004) \\
2001 Dec 10	 & 52254.295	 & 1 $\times$ 1920	   & $<$1	 & {\em Antu}	 & {\sc isaac}   & $J$  	 & Caballero et al. (2005) \\
2003 Jan 12	 & 52647.414	 & 12 $\times$ 1200	   & 4.2	 & INT  	 & {\sc wfc}	 & $I$  	 & Caballero et al. (2004) \\
2003 Oct 19	 & 52932.023	 & 1 $\times$ 3600	   & $<$1	 & 3.5\,m CA	 & {\sc o2k}	 & $H$  	 & Caballero et al. (2005) \\
2003 Oct 22	 & 52935.201	 & 1 $\times$ 600	   & $<$1	 & 3.5\,m CA	 & {\sc o2k}	 & $H$  	 & Caballero et al. (2005) \\
2003 Dec 22	 & 52995.596	 & 3 $\times$ 600	   & $<$1	 & TCS  	 & {\sc cain-2}  & $J$  	 & this paper \\
2003 Dec 22	 & 52996.086	 & 1 $\times$ 1200	   & $<$1	 & TCS  	 & {\sc cain-2}  & $H$  	 & this paper \\
2003 Dec 26	 & 53000.092	 & 3 $\times$ 120	   & $<$1	 & IAC-80	 & {\sc ccd}	 & $I$  	 & this paper \\
2003 Dec 28	 & 53002.080	 & 6 $\times$ 600	   & $\sim$1	 & IAC-80	 & {\sc ccd}	 & $I$  	 & this paper \\
2004 Jan 11	 & 53016.116	 & 10$\times$ 300	   & $\sim$1	 & OGS  	 & {\sc esaccd}  & $W$$^{\rm{a}}$& this paper \\
2004 Jan 14	 & 53019.061	 & 20 $\times$ 300	   & 1.9	 & OGS  	 & {\sc esaccd}  & $W$$^{\rm{a}}$& this paper \\
2004 Jan 24	 & 53028.859	 & 13 $\times$ 600	   & 2.2	 & IAC-80	 & {\sc ccd}	 & $I$  	 & this paper \\
2004 Jan 25	 & 53029.843	 & 14 $\times$ 600	   & 2.7	 & IAC-80	 & {\sc ccd}	 & $I$  	 & this paper \\
2004 Jan 30	 & 53034.879	 & 3 $\times$ 1200	   & $<$1	 & IAC-80	 & {\sc ccd}	 & $I$  	 & this paper \\
2004 Jan 30	 & 53034.854	 & 46 $\times$ 300	   & 4.0	 & OGS  	 & {\sc esaccd}  & $W$$^{\rm{a}}$& this paper \\
2004 Jan 31	 & 53035.887	 & 30 $\times$ 300	   & 2.7	 & OGS  	 & {\sc esaccd}  & $W$$^{\rm{a}}$& this paper \\
2004 Feb 1	 & 53036.833	 & 45 $\times$ 300	   & 6.3	 & OGS  	 & {\sc esaccd}  & $W$$^{\rm{a}}$& this paper \\
2004 Feb 2	 & 53037.875	 & 26 $\times$ 300	   & 4.6	 & OGS  	 & {\sc esaccd}  & $W$$^{\rm{a}}$& this paper \\
2004 Feb 3	 & 53038.948	 & 6 $\times$ 600	   & $\sim$1	 & IAC-80	 & {\sc ccd}	 & $I$  	 & this paper \\
           	\noalign{\smallskip}
            	\hline
         \end{tabular}
	\begin{list}{}{}
	\item [$^{\mathrm{a}}$] $W$: Optical images taken with no filter.
	\end{list}
\end{table*}

\subsubsection{$I$-band differential photometry} \label{I}

\begin{figure}
   \centering
   \includegraphics[width=0.48\textwidth]{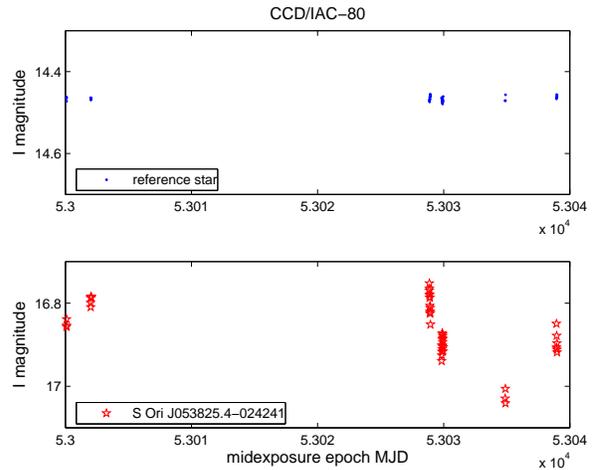}
      \caption{IAC-80 $I$-band light curves of S\,Ori
      J053825.4$-$024241 (bottom panel) and of a nearby reference star
      (top panel). Data have been obtained from Dec. 2003 through Feb. 2004. 
      Note that the vertical scale is the same for the two
      panels.  }
        \label{light_nIR+std_iac80_1}
   \end{figure}

Using the Thomson 1024 $\times$ 1024 {\sc ccd} camera (0.4325 arcsec
pixel$^{-1}$) at the IAC-80 telescope,
we have carried out time-series differential $I$-band photometry in a region of
6.5 $\times$ 6.5 arcmin$^2$ around S Ori J053825.4$-$024241. 
Our target was monitored during six nights. On two of these
nights, it was followed for more than two hours. A total of
45 images have been used to build the IAC-80 $I$-band light curve of
the object, which is shown in Fig. \ref{light_nIR+std_iac80_1}.  
The minimum variation amplitude detectable is 0.025\,mag.
The seeing ranged from 1.2 to 2.0 arcsec during the observations.  The
combined final image is depicted in Figure \ref{image_nIR_iac80}. The
photometric calibration of these data has been taken from B\'ejar et
al. (2004). 

\begin{figure}
   \centering
   \includegraphics[width=0.48\textwidth]{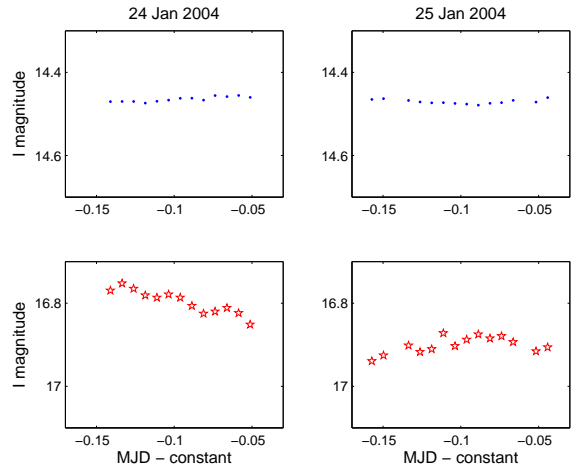}
      \caption{Details of the IAC-80 $I$-band light-curves of S\,Ori
      J053825.4$-$024241 (bottom panels) and of a nearby reference
      star (top panels). Our target was monitored for about 2.5\,h
      during MJD $\sim$ 53029 (left panels) and 53030 (right
      panels). It appears that a modulation of lower amplitude and
      shorter time scale is superimposed on a large photometric
      variability.}
         \label{light_nIR+std_iac80_2}
   \end{figure}

\begin{figure}
   \centering
   \includegraphics[width=0.48\textwidth]{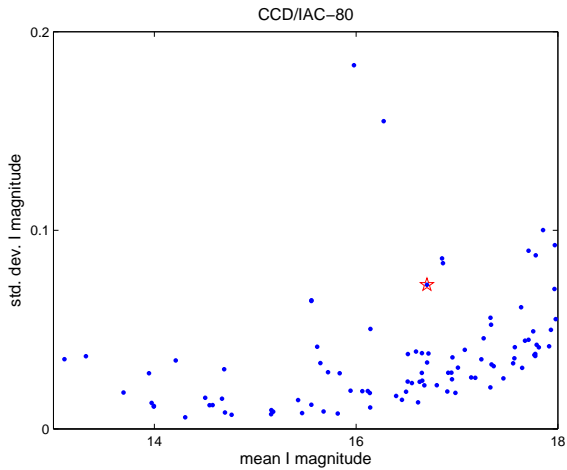}
      \caption{Standard deviation of the IAC-80 $I$-band differential
      photometry plotted against the average $I$ magnitude. The
      location of S\,Ori J053825.4$-$024241 is marked with a star
      symbol.}
         \label{light_iac80_variability}
   \end{figure}

Figure \ref{light_nIR+std_iac80_1} portrays the $I$-band light curve
of S\,Ori J053825.4$-$024241 and of one reference star. Error bars are
of the same size as the symbols in the plots. The data with the
longest temporal coverage are enlarged in
Figure \ref{light_nIR+std_iac80_2}. We have plotted the standard
deviation of the light curves of 97 sources in the field of view as a
function of $I$-band magnitude 
in Figure \ref{light_iac80_variability}.  S\,Ori J053825.4$-$024241
stands out from the trend defined by the great majority of the
sources\footnote{The most variable objects in this diagram are two sources
that are probably background large-amplitude eclipsing binaries that do not
belong to the cluster according to their $BRIJHK_{\rm s}$ magnitudes.
One of these binaries shows a light curve with primary and secondary minima of
similar amplitude ($\approx$ 0.6 mag) and a period of $\approx$ 5.5 h
(coordinates J2000: 05 38 19.50 --02 41 22.5).}.
S\,Ori J053825.4$-$024241 shows $I$-band variations of $\sim$0.05
and $\sim$0.25\,mag in timescales of hours and days, respectively.
We note the marked brightness decrease around MJD = 53035, which
lasted for about one week.


\subsubsection{White light differential photometry}

We have used the OGS telescope to photometrically monitor a large area
($\sim$0.5\,deg$^2$) of the $\sigma$ Orionis cluster. 
The mosaic detector of the {\sc esaccd} camera comprises four E2V
Technologies CCD42-40 devices (2048 $\times$ 2048 pixels,
0.62 arcsec pixel$^{-1}$), which provides a total field of view of about
0.5\,deg$^2$ in a single spot. We collected data from January 2004
through to the beginning of February 2004 on six different nights. Here
we will focus on the analysis of S\,Ori J053825.4$-$024241. 

We note that no filter was used when collecting the OGS data.  This
is indicated  by $W$ (for ``white light'') in the filter
column of Table \ref{log_phot}.  
Nevertheless, the detector is sensitive to
optical wavelengths. We have computed the effective wavelength of the
OGS observations for the case of S\,Ori J053825.4$-$024241 by
convolving the transmission function of the telescope, the response of
the E2V CCDs and the spectral energy distribution of an M6-type dwarf
(a significant fraction of the energy is released at red
wavelengths). Our computations yield an effective wavelength of
7800\,\AA, which is intermediate between the Johnson $R$ (6410\,\AA)
and $I$ bands (8500\,\AA) used at the IAC-80 telescope. 

\begin{figure}
   \centering
   \includegraphics[width=0.48\textwidth]{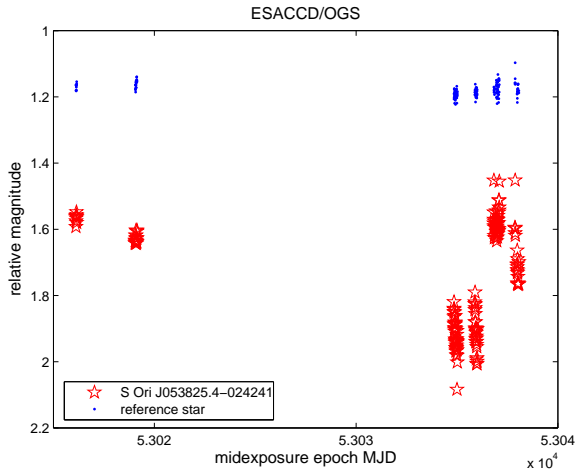}
      \caption{OGS optical light curve of S\,Ori J053825.4$-$024241
      (stars) and of a reference star (dots). Data have been obtained
      from Jan to Feb 2004.}
        \label{light_nIR+std_ogs_1}
   \end{figure}

\begin{figure}
   \centering
   \includegraphics[width=0.48\textwidth]{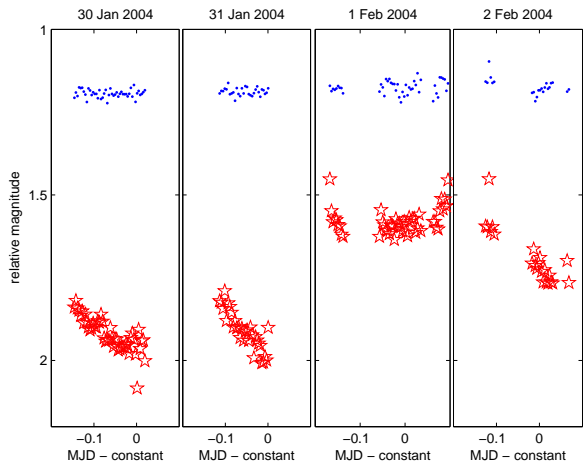}

      \caption{Details of the OGS light-curve of S\,Ori
      J053825.4$-$024241 (stars) and of a reference star (dots). Our
      target was monitored for 3--6\,h on each of these four observing
      nights.}

	 \label{light_nIR+std_ogs_2}
   \end{figure}

\begin{figure}
   \centering
   \includegraphics[width=0.48\textwidth]{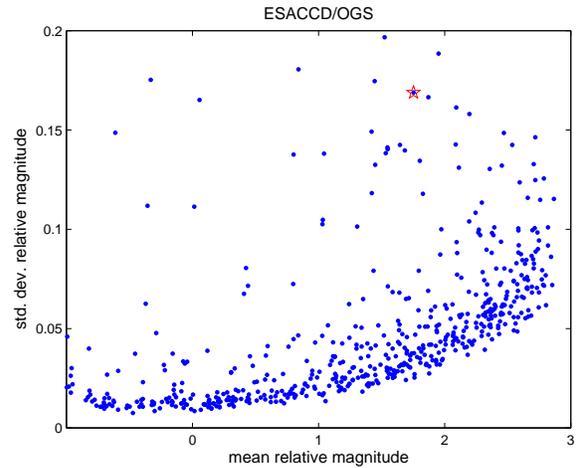}
      \caption{Standard deviation of the OGS differential
      photometry plotted against the average relative magnitude. The
      location of S\,Ori J053825.4$-$024241 is marked with a star
      symbol.}
	 \label{light_ogs_variability}
   \end{figure}

A total of 173
OGS images have been used to construct the light curve of our
target. We have not attempted to correct the light curve for
colour-dependent differential airmass extinction, as no
particular trend is observed between various sources and S\,Ori
J053825.4$-$024241 from night to night.
The minimum variation amplitude detectable from the OGS data is 0.020\,mag.
Figures \ref{light_nIR+std_ogs_1},
\ref{light_nIR+std_ogs_2} and \ref{light_ogs_variability} show the
complete OGS light curve of S\,Ori J053825.4$-$024241, the details of
four observing nights, and the standard deviation of the light curves
as a function of magnitude, respectively. S\,Ori J053825.4$-$024241
shows a marked variation amplitude at wavelengths shorter than
8000\,\AA: $\sim$0.2\,mag on timescales of hours, and 0.5\,mag from
night to night. The brightness decrease at around MJD = 53035 is also
observed in the OGS data.


\subsection{Near-infrared photometry}

We have obtained further near-infrared data ($J$ and $H$ bands) at different
epochs.  
All observations are summarized in Table \ref{log_phot}. 
The instruments used were the NICMOS-3 256 $\times$ 256 {\sc cain-2} 
camera (1.00 arcsec pixel$^{-1}$) at the TCS telescope,
{\sc omega-2000} at the 3.5 m CAHA telescope and {\sc isaac} at the UT1 {\it
Antu}.  
The $J$-band photometry obtained with the TCS in Jan 1999 was published in
B\'ejar et al. (2004).
The remaining near-infrared photometry of Table \ref{log_phot} is
unpublished and presented in this paper for the first time (see Table
\ref{data_phot}).
All near-infrared photometry has been referred to the 2MASS photometric
system (S\,Ori J053825.4$-$024241 was detected by 2MASS at MJD = 51116.323). 
We have carried out $J$- and $H$-band differential photometry of S\,Ori
J053825.4$-$024241 with respect to a certain number of reference stars within a
radius of 20 arcmin from our target. 
The minimum variation amplitudes that can be detected are 0.03\,mag in $J$ and
0.06\,mag in $H$.  
Our target has been observed in $K_{\rm s}$ only once by 2MASS.
Further discussion on the near-infrared variability of S\,Ori
J053825.4$-$024241 will be given in Section \ref{variability}.

\section{Discussion \label{discussion}}

\subsection{Mass estimate}
\label{mass_estimate}

Stauffer, Schultz \& Kirkpatrick (1998) and Mart\'\i n et al. (1998) determined that
the substellar mass limit lies at spectral type M6 in the Pleiades
cluster ($\sim$120\,Myr). State-of-the-art evolutionary models predict that
objects at the substellar boundary evolve at a roughly constant
effective temperature (within 200\,K) between a few million years and
the age of the Pleiades to cool down very fast afterwards. Thus, it
has been suggested that the star--brown dwarf boundary occurs at
M5--6 in $\sigma$ Orionis. The spectral type of S\,Ori
J053825.4$-$024241 is M\,6, i.e. close to the cluster substellar
limit. To constrain its mass with greater precision we  compare
the object luminosity to theoretical predictions. The luminosity of
S\,Ori J053825.4$-$024241 is derived from its $H$-band average
magnitude ($H$\,=\,14.18 $\pm$ 0.06) because, as we  discuss in
Section \ref{variability}, this object shows the lowest
photometric variability at these wavelengths. In addition, there is no
obvious excess emission at $H$ in our data. To transform this colour
magnitude into bolometric magnitude, we have used the bolometric
correction vs. spectral type relation of Golimowski et al. (2004)
and a distance of 360\,$\pm$\,70\,pc to the $\sigma$ Orionis cluster,
deriving $M_{\rm bol} =$ 9.0 $\pm$ 0.4. The large error bar associated
with this measurement is due to our poor knowledge of the cluster
distance.

\begin{figure}
   \centering
   \includegraphics[width=0.48\textwidth]{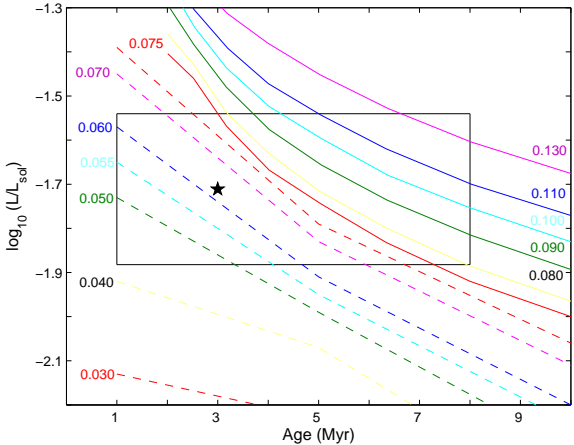}
      \caption{Models of luminosity evolution of low-mass stars and
      brown dwarfs from the Lyon group; full lines correspond to
      {\sc nextgen} models (Baraffe et al. 1998) and dashed lines to 
      {\sc dusty} models (Chabrier et al. 2000). 
      The most likely location of
      S\,Ori J053825.4$-$024241 is indicated by a star symbol. The
      box accounts for the uncertainty in the luminosity determination
      and the probable age of the cluster (1--8\,Myr). For ages below
      5\,Myr, S\,Ori J053825.4$-$024241 is very likely substellar. }
         \label{image_mass}
   \end{figure}

We show in Fig. \ref{image_mass} the luminosity evolution of objects
with masses between 0.03 and 0.13\,M$_\odot$ as provided by the Lyon
models (Baraffe et al. 1998, 2003; Chabrier et al. 2000).  The most
likely location of S\,Ori J053825.4$-$024241 in this diagram is
indicated by a star symbol at the age of 3\,Myr, the most probable age of
the cluster. The box around this
location accounts for the uncertainty in the luminosity determination
and the possible age interval (1--8\,Myr) of the $\sigma$ Orionis cluster.
From Fig. \ref{image_mass}, the mass of S\,Ori
J053825.4$-$024241 is estimated at 0.06$^{+0.07}_{-0.02}$\,M$_\odot$,
indicating that this object is probably substellar. 

\subsection{H$\alpha$ emission}

With a pEW of about $-$250\,\AA, S\,Ori J053825.4$-$024241 is one of the
strongest H$\alpha$ emitters in the $\sigma$ Orionis cluster.
The H$\alpha$ strength of our target is about five times larger than
that of the most active low-mass stars of the cluster, while only two
substellar objects, S\,Ori 55 and 71, show higher H$\alpha$ emission (Zapatero
Osorio et al. 2002b; Barrado y Navascu\'es et al. 2002). 
The marked persistence of the H$\alpha$ line of S\,Ori J053825.4$-$024241, which
is always seen in substantial emission in our data, suggests that its origin is
not due to a transient event.
As mentioned in Section \ref{optical_spectroscopy}, S\,Ori J053825.4$-$024241
shows broad H$\alpha$ wings (Fig. \ref{lris}). 
After correction for the contribution of the instrumental profile, the real
H$\alpha$ {\sc fwhm} turns out to be 230\,km\,s$^{-1}$ (or 220\,km\,s$^{-1}$ if
we consider the pedestal of the line at the 10\%~level of the peak flux). 
Such large velocity appears to be close to the break-up limit of S\,Ori
J053825.4$-$024241, so the H$\alpha$ broadening is probably not caused by
intrinsic rotation.
We interpret the persistent broad H$\alpha$ emission as due to hot mass infall
or mass ejection (e.g., White \& Basri 2003). 

\begin{figure}
   \centering
   \includegraphics[width=0.48\textwidth]{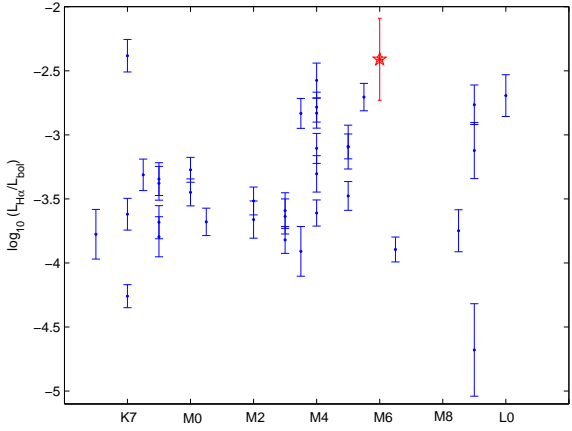}
      \caption{H$\alpha$ luminosity over bolometric luminosity is
      plotted against spectral type for $\sigma$ Orionis
      members. S\,Ori J053825.4$-$024241 (star symbol) exhibits a
      significant amount of H$\alpha$ emission in comparison to other
      cluster low-mass stars and brown dwarfs.}
	 \label{spec_LHaLbol}
   \end{figure}

To compare the relative strength of the H$\alpha$ emission of S\,Ori
J053825.4$-$024241 with those of other low-mass $\sigma$ Orionis 
members, we have evaluated the $L_{{\rm H}\alpha} / L_{\rm bol}$ ratio
of our target. This quantity is independent of the surface area and
represents the fraction of the total energy output in H$\alpha$. 
The bolometric luminosity was calculated using the $M_{\rm bol}$ given in Section
\ref{mass_estimate} and the solar bolometric magnitude of 4.74.
We followed Hodgkin et al. (1995) for obtaining $L_{{\rm H}\alpha}$ from
the width of the line, the $(R-I)_{\rm C}$ colour and the $V$-band
magnitude. The $(R-I)_{\rm C}$ colour is taken from B\'ejar et al. (2004),
who provide quasi-simultaneous photometry in the two bands. We have
estimated $V$ to be 21.0 $\pm$ 0.5\,mag by adopting the
$(V-I)_C$--spectral type relationship of Kirkpatrick et
al. (1994). The NOT and Keck H$\alpha$ measurements yield similar
values within the error bar, i.e. $\log{(L_{{\rm H}\alpha} / L_{\rm
bol})} = -$2.4 $\pm$ 0.3 (\footnote{If the object were brighter in $V$
than expected, e.g. $V$ = 20.0 $\pm$ 0.5, 
$\log{(L_{{\rm H}\alpha} / L_{\rm bol})}$ could be as high as $-$2.0 $\pm$
0.3.}).  
The error bar takes into account photometric
uncertainties and the error in the distance to the cluster. Figure
\ref{spec_LHaLbol} shows  $\log{(L_{{\rm H}\alpha} / L_{\rm bol})}$ as a
function of spectral type for cluster members. Except for our target,
data have been gathered from Zapatero Osorio et al. (2002a, 2002b) and
Barrado y Navascu\'es et al. (2002). S\,Ori J053825.4$-$024241
exhibits quite intense emission, which is comparable to that of active
stellar accretors.

\subsection{Other emission lines and radial velocity}

The ``photospheric'' radial velocity of S\,Ori J053825.4$-$024241
($v_h$\,=\,$+$32\,$\pm$\,13\,km\,s$^{-1}$) is consistent with 
membership of the $\sigma$ Orionis cluster.  
The cluster systemic velocity is in the range 27 to 38\,km\,s$^{-1}$ 
(Bohannan \& Garmany 1978; Garmany, Conti \& Massey 1980; Morrell \& Levato
1991). 
Zapatero Osorio et al. (2002a) and Kenyon et al. (2005) found mean
heliocentric velocities of 37\,$\pm$\,6 and 31.2\,$\pm$\,0.2\,km\,s$^{-1}$,
respectively, for low-mass members of the cluster. 

Emission lines, if originating outside the photosphere, may show a
different velocity. The measurements of H$\alpha$ and He {\sc i}
$\lambda$5876 are in agreement with the ``photospheric'' velocity of
S\,Ori J053825.4$-$024241 within the 1$\sigma$  uncertainty. However,
[O {\sc i}] $\lambda$6300 shows a moderate blue-shifted velocity of
about 2$\sigma$.  This may indicate the presence of low density gas
outflow (mass loss, wind), which is a process commonly found in Classical T
Tauri stars
with jets. The red component of this emission line might be hidden by
an opaque disc (Appenzeller, Jankovics \& \"Ostreicher 1984; Edwards et al.
1987). It has also been suggested (Corcoran \& Ray 1997) that low velocity
forbidden line emission (like the one we have measured in S\,Ori
J053825.4$-$024241) could be due to disc wind.

\subsection{Lithium and optical veiling}

Lithium at $\lambda$6707.8\,\AA~is not detected in any of our
spectroscopic data. 
From the observed spectra, we can impose upper limits on the line pEW of 
$+$0.7\,\AA~(NOT) and $+$0.3\,\AA~(Keck). Because of the very young
age of $\sigma$ Orionis, it is expected that all cluster stars and
brown dwarfs have depleted none of their initial lithium
abundance. 

   \begin{figure}
   \centering
   \includegraphics[width=0.48\textwidth]{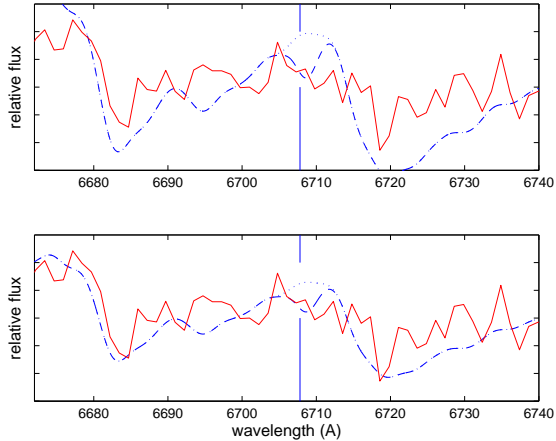}
      \caption{The Keck spectrum of S\,Ori J053825.4$-$024241 (solid line)
      is compared to ``unveiled'' synthetic spectra computed for log~$N({\rm Li})$ =
      3.0 (dashed line) and no lithium (dotted line) in the top panel.
      The bottom panel shows the
      comparison between S\,Ori J053825.4$-$024241 (solid line) and a ``veiled''
      ($r$ = 0.5) theoretical spectrum with log~$N({\rm Li})$ = 3.0 (dashed line). 
      The location of the lithium feature is indicated by a vertical line.}   
           \label{sed_yp} 
   \end{figure}

We have computed a few synthetic spectra around lithium using the code of
Pavlenko, Zapatero Osorio \& Rebolo (2000), 
and $T_{\rm eff}$\,=\,2900\,K, $\log{g}$\,=\,4.0 and
solar metallicity, which are the expected photospheric parameters of an M6-type
dwarf in $\sigma$ Orionis. The theoretical spectra have been degraded to match
the resolution of 4\,\AA. Computations have been carried out considering two
different lithium abundances (log\,$N({\rm Li})$\,=\,3.0, i.e. no depletion, and
complete depletion), and the presence of veiling. 
S\,Ori J053825.4$-$024241 shows a very strong H$\alpha$ emission suggestive
of intense accretion; 
furthermore, the $R-I$ colour and the titanium oxide bands of this object appear
bluer and weaker, respectively, than expected for M6 dwarfs, possibly
indicating the presence of some blue continuum excess. 
For our computations, we have adopted a flat energy distribution in the
wavelength interval 6672--6740\,\AA~for the flux continuum excess. 
Our results are depicted in Fig. \ref{sed_yp}, where
the top panel shows the comparison between the observed Keck spectrum of S\,Ori
J053825.4$-$024241 and the ``unveiled'' synthetic spectra, and the bottom panel
displays S\,Ori J053825.4$-$024241 and the ``veiled'' computations. It is
apparent from the figure that the TiO features of S\,Ori J053825.4$-$024241 are
better reproduced by the spectral synthesis including some veling.
An amount of $r$\,=\,0.5, where $r = F_{\rm excess}/F_{\rm phot}$,
reasonably matches the observations.  
The lithium absorption feature is detectable in the unveiled, 
lithium-rich, 4\,\AA-resolution theoretical spectrum (pEW $\approx$
0.3\,\AA),  
whereas it becomes rather weak in the veiled spectrum. Therefore, the 
non-detection of lithium in our data is still consistent with complete 
preservation. Higher-resolution spectra are needed to study lithium in 
S\,Ori J053825.4$-$024241.

\subsection{Photometric variability and the Classical T Tauri scenario
\label{variability}} 

Based on our optical data, we conclude that S\,Ori J053825.4$-$024241
presents the largest variability among very low-mass stars and brown
dwarfs so far identified in the $\sigma$ Orionis cluster
The amplitude of the variations in the $I$-band (8500\,\AA) are
$\sim$0.05\,mag in timescales of hours and $\sim$0.25\,mag on 
timescales of days. The variability is even larger at shorter
wavelengths. 
These conclusions strongly support the Classical T Tauri (CTT)
picture. 
As noted by Appenzeller \& Mundt (1989), all carefully monitored CTT stars have
turned out to be variables.  
We have searched for a periodic signal in the IAC80 and OGS data of
S\,Ori J053825.4$-$024241 by calculating the power spectrum as described in
Caballero et al. (2004).
We have found no significant peaks in the periodograms.

   \begin{figure}
   \centering
   \includegraphics[width=0.48\textwidth]{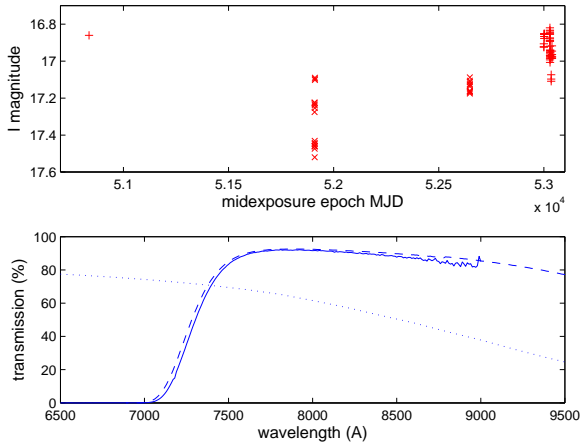}
      \caption{The long-term $I$-band light-curve of S\,Ori
      J053826.1$-$024041 is depicted in the top panel. Data, which
      include the INT (crosses) and IAC-80 (pluses) photometry, span a
      time coverage of six years. The bottom panel displays the
      transmission curves of the IAC-80 (dashed line) and the INT
      (RGO-I, full line) $I$-band filters and of one of the CCDs
      (dotted line) of the INT camera. }
         \label{light_optical_long}
   \end{figure}

   \begin{figure}
   \centering
   \includegraphics[width=0.48\textwidth]{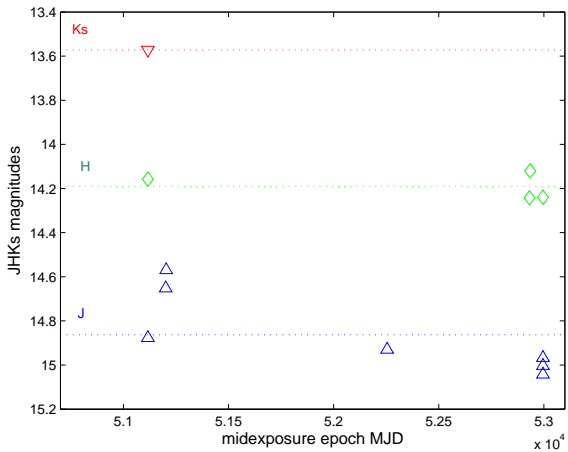}
      \caption{$JHK_{\rm s}$-band light-curves of S\,Ori
      J053825.4$-$024241 
      ($J$: $\bigtriangleup$; $H$: $\diamond$; 
      $K_{\rm s}$: $\bigtriangledown$). 
      The dotted lines stand for the average
      magnitude in each filter.}
         \label{light_JHKs}
   \end{figure}

To study the long-term optical photometric variability of S\,Ori
J053825.4$-$024241 we have compared our data to previously published
data obtained at different epochs and using similar instrumentation. 
These data and the new photometry presented here span a total of 6 years of
observations.  
The $I$-band filters used at the IAC-80 and INT are very similar as inferred
from their transmission curves shown in the bottom panel of Figure
\ref{light_optical_long}.  
For direct comparison, all data have been referred to the
same photometric calibration as in B\'ejar et al. (2004).
The comparison of the light curves of the reference stars in common between
the IAC-80 and INT datasets yields a dispersion of 0.025\,mag.  
The complete set of $I$-band observations is depicted in the top panel of
Figure \ref{light_optical_long}. S\,Ori J053825.4$-$024241 also presents
long-term photometric varibility in the optical with an amplitude of
0.7\,mag. This is larger than the short-term variation
amplitude. Further data are needed to investigate whether the optical 
light curve of S\,Ori J053825.4$-$024241 consists of a long, very large
amplitude variation on to which variations of smaller amplitudes and
shorter timescales are superimposed (see Scholz \& Eisl\"offel 2004).
We provide in Table \ref{data_phot} the intervals of the photometric
magnitudes that we have measured in the optical and near-infrared.

Figure \ref{light_JHKs} shows all near-infrared photometry of S\,Ori
J053825.4$-$024241 available to date as a function of the epoch of the
observations. 
The $H$-band photometry appears to be rather stable within 1$\sigma$
the uncertainty.  
At $J$, our target shows a large scatter, suggesting variability. 
The amplitude of the variations is about 0.5\,mag, which compares with the
optical variability. 
The photometric variability of CTT stars decreases towards longer wavelengths
because the origin of the variations probably lie in hot and/or cold
photospheric spots covering a significant percentage of the object surface. 
It is also believed that part of the variability is due to occultations by
inhomogeneities of the inner disc regions, which show less dependence on
wavelength.  
This picture might be applicable to S\,Ori J053825.4$-$024241 to explain its
peculiar optical and near-infrared light-curves. We note that S\,Ori
J053825.4$-$024241 presents a clear flux excess at 2.2\,$\mu$m
(B\'ejar et al. 2004; Caballero et al. 2004), which may be associated with
the presence of a surrounding disc.

\begin{table}
    \centering
    	\caption[]{Photometric data of S\,Ori J053825.4$-$024241}
        \label{data_phot}
        \begin{tabular}{lcr}
            	\hline
            	\noalign{\smallskip}
Filter			& Magnitude             & Source \\		
            	\noalign{\smallskip}
            	\hline
            	\noalign{\smallskip}
$B^{\rm{a}}$		& $\sim$ 20.9		& USNO-B1 \\ 
$R^{\rm{a}}$		& 18.66 $\pm$ 0.05	& B\'ejar et al. (2004) \\ 
$I$			& 16.82--17.52		& this paper \\ 
$J$			& 14.57--15.04		& this paper \\ 
$H$			& 14.18 $\pm$ 0.06	& this paper \\ 
$K_{\rm s}^{\rm{a}}$ 	& 13.57 $\pm$ 0.03 	& 2MASS \\
           	\noalign{\smallskip}
            	\hline
         \end{tabular}
	\begin{list}{}{}
	\item [$^{\mathrm{a}}$] A single measurement.
	\end{list}
\end{table}

Interestingly, S\,Ori J053825.4$-$024241 is detected in the USNO-B1
blue plates (Monet et al. 2003), from which we have obtained the blue
photographic magnitude. This magnitude has been converted into the
standard $B$ magnitude by applying the conversion equation of Zombeck
(1990). We provide our measurement in Table \ref{data_phot}. From the
$B$ magnitude and the typical colours of M5--7 dwarfs, it is also
apparent that S\,Ori J053825.4$-$024241 presents a flux excess at blue
wavelengths. This result is consistent with the presence of veiling
as discussed in the previous section. 
We note that on some occasions, when the
accretion activity is large, S\,Ori J053825.4$-$024241 may show up
bluer in the optical than other cluster members of similar
types. This might explain why this object was not selected
as a cluster member candidate from previous $RI$ surveys in the region.

\section{Summary and conclusions}

We have obtained low-resolution optical spectroscopy ($R$ $\sim$ 600)
in the wavelength interval 5300--9600\,\AA~of four low-mass candidate
members of the $\sigma$ Orionis cluster (360 $\pm$ 70\,pc,
1--8\,Myr). These objects were selected from the photometric surveys
by B\'ejar et al. (2004) and Caballero et al. (2004). We have
determined spectral types between M5 and M6 for these sources, which
are consistent with our expectations for true cluster members. 
Furthermore, all of them show spectroscopic features typical
of low-gravity atmospheres, supporting their membership of the
cluster.

One object, namely S\,Ori J053825.4$-$024241 (M\,6.0$\pm$1.0), was known to
be variable in the $I$ band (Caballero et al. 2004) and to show
slightly bluer $R-I$ and redder $J-K_{\rm s}$ colours than other
$\sigma$ Orionis members of similar magnitude (B\'ejar et
al. 2004). From our low-resolution spectroscopic data, S\,Ori
J053825.4$-$024241 presents a strong H$\alpha$ emission (one of the
largest among known low-mass cluster members), which, in addition to
the near-infrared flux excess, suggests that this object is undergoing
accretion processes from a surrounding disc.

To confirm the possible T Tauri nature of S\,Ori J053825.4$-$024241,
we have obtained follow-up higher resolution spectra ($R$ $\sim$
1700) in the wavelength interval 5600--8100\,\AA~and have
photometrically monitored the object in the $I$, $J$ and $H$ bands 
and in white light. 
S\,Ori J053825.4$-$024241 shows persistent, strong 
H$\alpha$ emission. In addition, other forbidden and permitted
emission lines have been detected, e.g. $[$O{\sc i}$]$
$\lambda$6300.3 and He {\sc i} $\lambda$5875.8. The pseudo-equivalent
widths of all these emission lines are consistent with the
measurements of known stellar accretors. Furthermore, H$\alpha$ is
found to be broadened with widths typical of T Tauri stars, which
indicates the presence of gas infall at high velocities. 
From the comparison of our spectroscopic data to theoretical spectra, we
infer that some continuum flux excess is present at short wavelengths.
We quantify this ``veiling'' to be around 0.5 in the Li {\sc i}
$\lambda$6707.8 region. 
S\,Ori J053825.4$-$024241 has a radial velocity consistent with membership 
of the $\sigma$ Orionis cluster.

S\,Ori J053825.4$-$024241 is confirmed to be an irregular variable object. 
Variations, which do not show any obvious modulation pattern,
are significant (0.05--0.7\,mag) from blue wavelengths up to the
$J$ band. Variability is found at all timescales. Our $H$-band
photometry appears rather stable in time, suggesting little
variability at wavelengths greater than 1.2\,$\mu$m. Both the
photometric and spectroscopic properties of S\,Ori J053825.4$-$024241
are consistent with the T Tauri scenario.
Assuming that S\,Ori J053825.4$-$024241 is a member of the $\sigma$
Orionis cluster, we have estimated its mass at
0.06$^{+0.07}_{-0.02}$\,M$_\odot$ after comparison with
state-of-the-art evolutionary models. Based on our data, we conclude
that S\,Ori J053825.4$-$024241 is probably a brown dwarf undergoing
processes similar to those of T Tauri stars.

\begin{acknowledgements}
  	We would like to thank Terry Mahoney for revising the English of the 
 	manuscript and the anonymous referee for 
	comments and suggestions that improved the article. 
	JAC thanks to staff at the Observatorio del Teide for performing the
	Observaciones de Tiempo de Servicio and to David Barrado y Navascu\'es,
	Gabriel Bihain, Jorge Sanz Forcada and Mar\'{\i}a del Mar Sierra for
	helpful comments. 
	Partial financial support was provided by the Spanish Ministerio de
	Ciencia y Tecnolog\'{\i}a, project AYA2001-1657 of the Plan Nacional de
	Astronom\'{\i}a y Astrof\'{\i}sica, and Ministerio de Educaci\'on y
	Ciencia, projects AYA2001-1657 and AYA2003-05355.
	YP's studies are partially supported by Royal Society and Leverhulme
	Trust grants.
	Research presented herein was partially funded by NSF research grant AST
	02-05862. 
	Acknowledgements of the use of telescopes, instruments, catalogues
	and software can be found in the electronic version.
	 
	{\Large ONLY IN THE ELECTRONIC VERSION:}
	
	The Telescopio IAC-80 and the Telescopio Carlos S\'anchez are
	operated on the island of Tenerife by the Instituto de Astrof\'{\i}sica
	de Canarias in the Spanish Observatorio del Teide of the Instituto de 
	Astrof\'{\i}sica de Canarias.	
     	Based on observations made with the European Space Agency Orbital Ground
	Station telescope operated on the island of Tenerife by the Instituto de
   	Astrof\'{\i}sica de Canarias in the Spanish Observatorio del Teide of
   	the Instituto de Astrofísica de Canarias.
	Based on observations made with the Nordic Optical Telescope, operated
	on the island of La Palma jointly by Denmark, Finland, Iceland,
	Norway, and Sweden, in the Spanish Observatorio del Roque de los
	Muchachos of the Instituto de Astrof\'{\i}sica de Canarias. 
	The data presented here have been taken using ALFOSC, which is owned by
	the Instituto de Astrof\'{\i}sica de Andaluc\'{\i}a (IAA) and operated
	at the Nordic Optical Telescope under agreement between IAA and the
	NBIfAFG of the Astronomical Observatory of Copenhagen.  	
	Some of the data presented herein were obtained at the W. M. Keck
	Observatory, which is operated as a scientific partnership among the
	California Institute of Technology, the University of California and the
	National Aeronautics and Space Administration. 
	The Observatory was made possible by the generous financial support of
	the W. M. Keck Foundation.  
	The authors wish to recognize and acknowledge the very significant
	cultural r\^ole and reverence that the summit of Mauna Kea has always
	had within the indigenous Hawaiian community.  
	We were most fortunate to have had the opportunity to conduct observations
	from this mountain. 
	This research has made use of the SIMBAD database, operated at CDS,
 	Strasbourg, France.  
	This publication makes use of data products from the Two Micron All Sky
	Survey, which is a joint project of the University of Massachusetts and
	the Infrared Processing and Analysis Center/California Institute of
 	Technology, funded by the National Aeronautics and Space
 	Administration and the National Science Foundation.
	The Guide Star Catalogue-II is a joint project of the Space Telescope
	Science Institute and the Osservatorio Astronomico di Torino. Space
	Telescope Science Institute is operated by the Association of
	Universities for Research in Astronomy, for the National Aeronautics and
	Space Administration under contract NAS5-26555. The participation of the
	Osservatorio Astronomico di Torino is supported by the Italian Council
	for Research in Astronomy. Additional support is provided by  European
	Southern Observatory, Space Telescope European Coordinating Facility,
	the International GEMINI project and the European Space Agency
	Astrophysics Division. 
	{\sc iraf} is distributed by National Optical Astronomy Observatories,
	which are operated by the Association of Universities for Research in
	Astronomy, Inc., under cooperative agreement with the National Science
	Foundation.

\end{acknowledgements}

\end{document}